\documentclass[aps,preprint,showpacs,groupedaddress]{revtex4}
\usepackage{graphics}
\usepackage{graphicx}
\usepackage{amsmath}
\usepackage{amssymb}
\usepackage{bm}
\usepackage{dcolumn}
\begin{document}
\preprint{version 1.0}
 \title[]{
Wetting phenomenon in the  liquid-vapor phase coexistence of a
partially miscible Lennard-Jones binary mixtures} 
\author{Enrique D\'\i az-Herrera$^a$,  Jos\'e A. Moreno-Razo$^a$ and 
Guillermo Ram\'\i rez-Santiago$^b$}
\affiliation{$^a$Departamento de F\'\i sica, 
Universidad Aut\'onoma Metropolitana-Iztapalapa,
Apdo. Postal 55-534, M\'exico 09340, D.F., MEXICO\\
$^b$Instituto de F\'\i sica, Universidad Nacional Aut\'onoma de
M\'exico, Apdo. Postal 20-364 M\'exico 01000, D. F. MEXICO}
\begin{abstract}
We have carried out extensive equilibrium molecular dynamics (MD)
simulations to study the structure and the interfacial properties 
in the liquid-vapor (LV) phase coexistence of partially miscible 
binary Lennard-Jones (LJ) mixtures. By analyzing the structural 
properties as a function of the miscibility parameter, $ \alpha $,
we found that at relatively low temperatures the system separates 
forming  a liquid A-liquid B interface in coexistence with the vapor 
phase. At higher temperatures and, $ 0<\alpha\leq 0.5 $, we found a 
temperature range,  $T^{*}_{w}(\alpha ) \leq T^{*} < T^{*}_{\mbox cons}(\alpha )$, 
where the liquid phases are wet by the vapor phase.  Here, $ T^{*}_{w}(\alpha) $ 
represents the wetting transition temperature (WTT) and 
$T^{*}_{\mbox cons}(\alpha )$ is the consolute temperature of the mixture. 
However, for $ 0.5< \alpha < 1$, no wetting phenomenon occurs. For the 
particular value, $ \alpha=0.25 $, we analyzed quantitatively the $T^{*}$ 
versus $\rho^{*}$, and $ P^{*} $ versus $ T^{*} $ phase diagrams and found,
$ T^{*}_{c}\simeq 1.25 $,  and $T^{*}_{\mbox cons}\simeq1.25$. 
We also studied quantitatively, as a function of temperature, the surface 
tension and the adsorption of  molecules at the liquid-liquid interface. 
It was found that the adsorption shows a jump from a finite negative value 
up to minus infinity, when the vapor wets the liquid phases, suggesting that
the wetting transition (WT) is of first order. The calculated phase diagram 
together with the wetting phenomenon strongly suggest the existence of a 
tricritical point. These results agree well with some experiments carried out 
in fluid binary mixtures.
\end{abstract}
\maketitle
\section{Introduction}
\label{introduction}
Wetting phenomena occurs very often in daily life and has a strong 
interdisciplinary character. It is of great 
relevance for fundamental areas of knowledge such as 
physics~\cite{koshevnikov97,physics}, 
chemistry~\cite{chemistry}, biology~\cite{biology0}, 
and several other applied  sciences~\cite{metallurgy} as well 
as technology~\cite{technology0}. 
The occurrence of wetting phenomenon is usually associated with the
existence of  three phases, at least one of which is liquid and no 
more than one phase is solid. In most  practical situations the 
solid  phase is wet by the liquid phase and the  disfavored phase 
is the vapor. This is expressed in terms of surface free 
energies as, $\gamma_{SL} < \gamma_{SV}$. 
Although the vapor phase is disfavored, a solid-vapor coexistence can be 
possible and in such case it is said that the liquid {\it "partially wets"} 
the solid. However, as the liquid phase is further adsorbed 
by the solid, it is possible that the contact of the vapor with the solid be 
excluded by the creation of a liquid layer between them. In such a situation 
one says that the liquid {\it ``completely wets"} the solid. The surface free 
energies involved in the wetting process are related by Young's rule, 
$\gamma_{SV}=\gamma_{SL} + \gamma_{LV}\cos\theta$, with $\theta$  the
contact angle. For partial wetting, $0<\theta< 90^{0}$, and for complete 
wetting, $\theta=0$.  
Usually the presence of a wall in most of the theoretical as
well as experimental studies complicates a detailed and precise
investigation of the interfacial structures. This is due to the fact
that the atomic interactions between the wall and the fluid components 
play an important role in the wetting phenomena.  
Fortunately, binary liquid mixtures offer a great
opportunity to investigate in detail wetting phenomena since they involve
only fluid phases in coexistence. 
An important number of experimental studies of interfacial wetting have
studied only a small subset of possible binary 
mixtures~\cite{koshevnikov97,kwon82,houessou85,kahlweit88,bonn92,kellay93}.
Wetting and prewetting phenomena in which one of the phases is solid has 
also been studied by means of Monte Carlo simulations using a 3D lattice-gas 
model~\cite{nicolaides89}.  A 12-6 LJ potential modeling argon and a 9-3 LJ
potential modeling a CO$_{2}$ covered solid wall have been chosen to study the 
wetting transitions by means of Monte Carlo simulations~\cite{finn89}.
However, to the best of our knowledge, there are not numerical 
simulations that explicitly study the interfacial behavior and 
the associated wetting phenomena in fluid mixtures in absence of a wall.  
In this paper we apply a well established methodology using
MD simulations~\cite{chapela,alejandre,mecke,nosotros,holc93} 
to study interfacial properties and surface phenomena 
in a partially miscible fluid mixture.
We consider a model binary LJ fluid mixture in which the
attractive part of the A-B interactions is weaker than the
A-A and B-B interactions. By studying the properties of the 
density profiles of the fluid phases as a function of temperature
and miscibility parameter,  we show clear evidence that  the vapor 
phase spontaneously excludes the liquid A-liquid B interface at and above
the wetting transition temperature, $ T^{*}_{\mbox w} $. 
That is, the vapor phase fully wets the liquid A-liquid B 
phases. This phenomenon occurs when the miscibility parameter 
$ \alpha $ is in the range  $ 0 < \alpha \leq 0.5 $ and in the temperature 
region  $ T_{W}^{*}(\alpha) < T^{*} < T_{\mbox{cons}}^{*}(\alpha)$. 
Here, $T^{*}=(k_{B}T)/\epsilon$, is the reduced temperature,
$T_{\rm w}^{*}(\alpha) $, is the wetting transition temperature (WTT), and
$T_{\mbox{cons}}^{*}(\alpha) $, is the consolute temperature of the mixture.
By analyzing the structural properties of the mixture as a function of 
$ \alpha $ we estimate the wetting phase diagram  $ T_{\rm w}^{*}(\alpha)$. 
A further quantitative analysis, for $ \alpha=0.25 $, of the surface free 
energy $ \gamma(T^{*}) $ and the adsorption, $\Gamma (T^{*})$, at the LL 
interface indicates that the wetting transition is of first order.

The layout of the remainder of this paper is as follows. 
In section~\ref{model} we introduce the model of the symmetric 
binary mixture,  in section~\ref{details} we explain the  details 
of the simulations, then in section~\ref{results} we present and 
discuss the representative results of the extensive MD simulations. 
Finally, we end with the conclusions in section~\ref{conclusions}.

\section{The model} 
\label{model}  
The model binary mixture studied in this paper consists of fluids 
A and B made up of spherical molecules of the same size, 
$ \sigma_{AA}=\sigma_{BB}$, and at concentrations of 50\% each.
The interaction between molecules of the same type is 
through the classical 12-6 LJ potential. However, 
the interaction between molecules of fluids A and B is given 
by the LJ potential,
\begin{equation}
\label{potential}
u_{ij}(r) =
\left\{ \begin{array}{ll}
4\epsilon_{ij}\left[ \left( \frac{\sigma_{ij}}{r_{ij}} \right)^{12} 
- \left( \frac{\sigma_{ij}}{r_{ij}} \right)^6 \right], 
& \mbox{if} \>\> r\leq R_{c}=3\sigma_{AA}, \\
  0, 
& \mbox{if}  \>\>r>R_{c}=3\sigma_{AA},
\end{array} \right.
\end{equation}
where the mixing rule is defined by 
\begin{eqnarray}
\sigma_{AB}&=&\frac{1}{2}(\sigma_{AA}+\sigma_{BB}),\cr
\epsilon_{AB}&=& \alpha_{AB}\epsilon_{AA},
\label{mix-rule}
\end{eqnarray} 
with  $\epsilon_{AA} = \epsilon_{BB}$, and  $\alpha_{AB}$ the 
parameter that controls the miscibility of the two fluids. 
For the sake of simplicity from now on we will use 
the short notation, $ \alpha_{AB} =\alpha$. Notice that 
when, $ \alpha=0 $, we obtain two independent single LJ fluids 
while in the opposite case, $ \alpha=1 $, the system reduces to 
a single LJ fluid. By choosing, $0 < \alpha < 1$, the attractive part 
of the A-B interactions becomes weaker than that of the AA and BB 
interactions, and then, the liquid phases are immiscible in a wide 
range of temperatures. Thus, one can obtain the coexistence of three 
fluid phases: liquid A-liquid B and the vapor. 
\section{Details of the simulations}
\label{details}  
\begin{figure}
\begin{center}
\includegraphics[height = 5.0in, width=5.0in, keepaspectratio=true]
{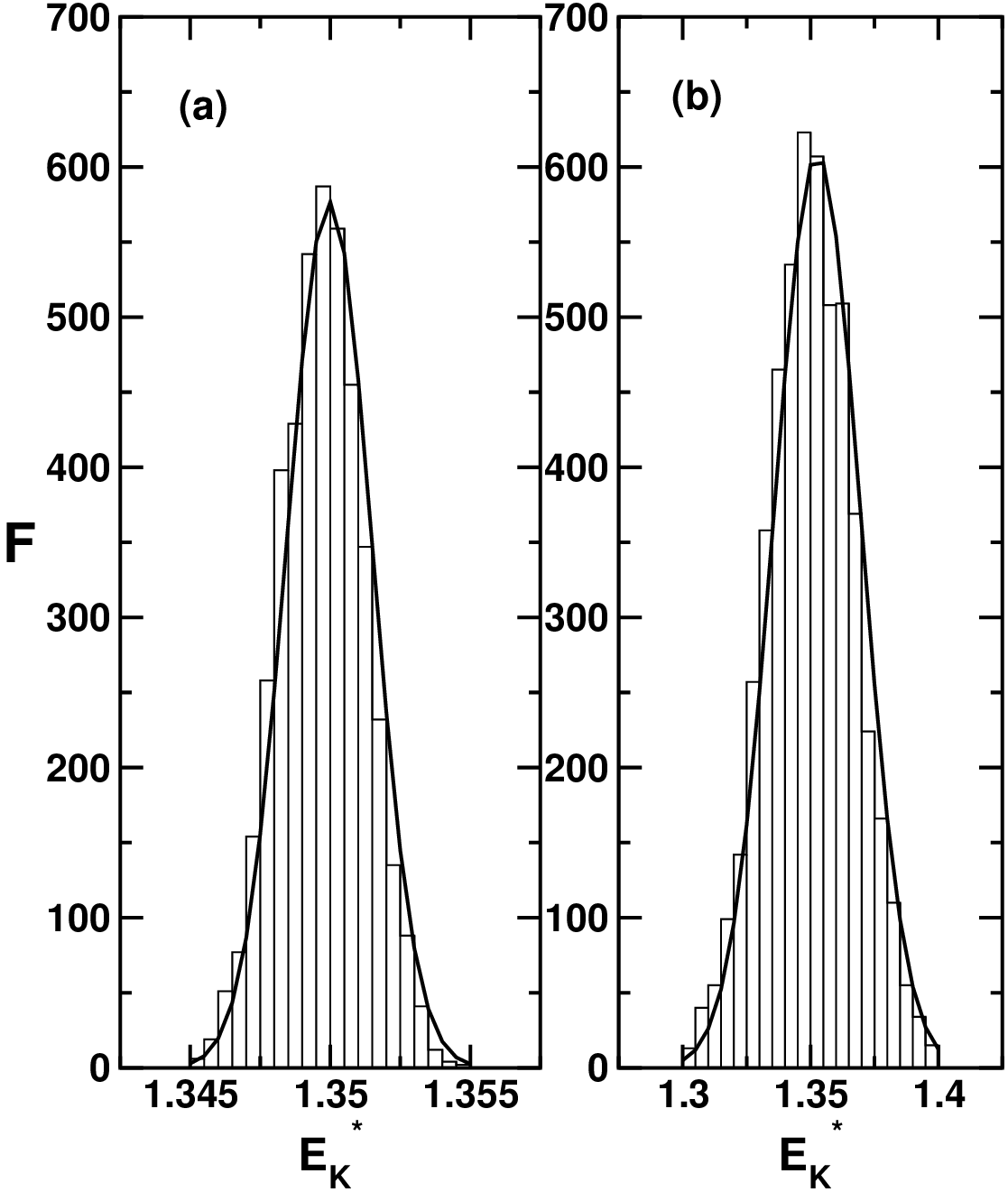}
\caption{Frequency versus reduced kinetic energy
(Kinetic energy distribution) for a binary mixture
with $ \alpha=0.25$ and  $ N=4096 $ particles at $ T^{*}= 0.9$.
The vertical axis should be multiplied by a factor of $ 10^{3} $.
(a) Results using an isokinetic thermostat and, (b) results
applying a Nos\'e-Hoover thermostat. In each case the solid line 
represents the best fit to a  Maxwell-Boltzmann distribution.}
\label{veloc-dist}
\end{center}
\end{figure}

We have carried out extensive MD simulations to investigate the
structural properties of this model binary mixture as a function of 
$ \alpha $, varying this quantity in steps, $ \delta\alpha =0.05$, in the interval
$ 0.2 \leq  \alpha \leq 0.5$.  For $ \alpha=0.25 $ we quantitatively related the
interfacial properties with the corresponding phase diagram properties. 
In all the simulations we applied  periodic boundary conditions along 
the $x,y$ and $z$  directions. At the lowest temperature, $T^{*}=0.65$, 
the simulations were initiated from a configuration where the molecules of type A 
and B  form two contiguous FCC crystals. At higher temperatures we take as 
the initial configuration the final configuration of the previous temperature. 
The initial velocities of the molecules were 
assigned from a Maxwell-Boltzmann distribution. The equations of motion 
were integrated using a leap frog algorithm with a time step size, 
$\delta t^{*}=0.005$. This corresponds to $1.1\times 10^{-5}$ 
nanoseconds in the scale of argon. At each time step iteration we monitor 
the temperature of the system, by means of the equipartition theorem,  
and rescale the linear momentum of the molecules to keep the temperature 
constant. This method of rescaling the linear momenta  is known as  the 
isokinetic thermostat. To check that this isokinetic thermostat produces 
meaningful results we have also carried out some MD simulations using the 
Nos\'e-Hoover thermostat. In figure~\ref{veloc-dist} we show the kinetic 
energy distributions as the dynamics of the system evolves at, $ T^{*}=0.90 $, 
applying both methods. 
As one would expect both thermostats yield a Gaussian distribution of kinetic
energies. The only difference is the width of the distributions. 
The isokinetic thermostat produced a distribution with a variance that is one 
order of magnitude smaller than the  variance of the distribution obtained 
with the  Nos\'e-Hoover thermostat. As a further check of the isokinetic thermostat
we also calculated and monitored some thermodynamic and surface quantities 
applying both thermostats. The results of this comparison test showed full consistency.   
Since MD simulations using the 
Nos\'e-Hoover thermostat are computationally more demanding we used the 
isokinetic thermostat in all the simulations reported in this paper.
Furthermore, to check the stability of the interfaces as well as the 
distribution of the species, we simulated the system for as long as
55 ns in the scale of Argon. Thermodynamic quantities and interfacial
properties of interest were measured by averaging over the last 
million of time-step iterations. To minimize correlations between 
measurements we calculated thermodynamic, structural and surface
quantities every 50 time steps.
We also investigated the role of finite 
size effects for the value of the miscibility parameter, $ \alpha=0.25 $. 
To this end  we carried out MD simulations with three system 
sizes, N=$1728,\>4096$  and 6144 molecules. We found that for all the 
quantities studied here, simulations with $ N=4096 $ provided reliable 
results. Therefore, most of the simulations were carried out with $N=4096$ 
molecules. The discussion of these results will be presented, where 
appropriate, in the next section. 
On the other hand, interfacial properties are sensitive to the cross section 
area of the simulational box that is parallel to the interfaces, as discussed 
in previous MD simulations of the LV interface of a single LJ 
fluid~\cite{holc93,li95}.  These authors arrived  to the conclusion that a 
reliable value of the cross section area of the computational box  should 
be at least $(8\sigma)^2$.  So, to be on the safe side, in the present simulations  
we have considered a computational box with a cross section area, 
$Lx \times Ly= (9\sigma_{AA})^{2}$. 
The length,  $L_{z}$, of the simulational parallelepiped was adjusted 
such that the average density of the system laid somewhere inside 
the LV coexistence curve. In this way one readily gets the liquid-vapor 
phase coexistence. The average densities of the simulated systems were in the 
range $ 0.2 \leq \rho^{*} \leq 0.4 $, where the reduced density defined as, 
$ \rho^{*} =\rho\sigma_{\rm AA}^{3}$. In the following section we present, 
analyze and discuss the results of the thermodynamic and interfacial 
properties calculated from our MD simulations.
\section{Results and discussion}
\label{results}
\subsection{Structural properties and phase diagrams}
\label{structure}
We performed extensive MD simulations for mixtures with
$0.2 \leq \alpha \leq 0.65 $ changing this parameter in steps of,
$ \delta\alpha\>=0.05 $. We studied the density profiles, $ \rho^{*}(z) $, 
of the liquid-vapor coexistence in different temperature regions.
From an analysis of $ \rho^{*}(z) $, as function of $ T^{*} $, and, 
$ \alpha $, we estimated the $ T_{\rm W} $ versus $ \alpha $ 
phase diagram. In what follows we present some representative results 
for, $ \rho^{*}(z) $, when $ \alpha=0.25 $ and 0.30.
Then, for the particular value, $ \alpha=0.25 $, we present a quantitative 
analysis of the $ T^{*} $ versus $\rho^{*}$, and $ T^{*} $ versus $ P^{*} $ 
phase diagrams. We  also give a brief description  of the procedure we followed 
to locate the coexistence courve and the --$ \lambda $ line-- mixing-demixing 
line.

\begin{figure}
\begin{center}
\includegraphics[height = 5.0in, width=5.0in, keepaspectratio=true]
{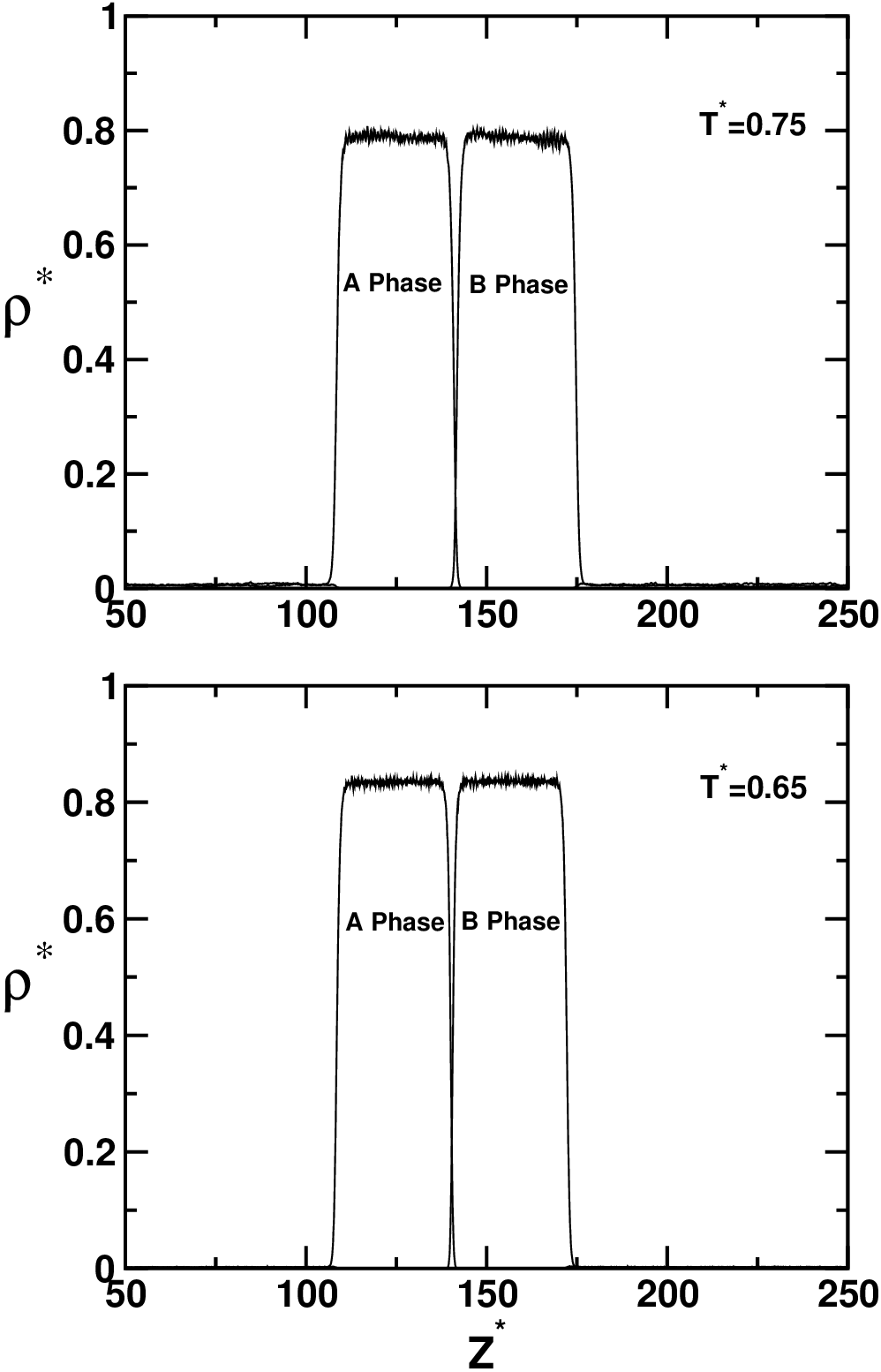}
\caption{Liquid-vapor-liquid reduced density profiles of the mixture
for $ \alpha=0.25 $  and $N=4096$. Note that at these 
relatively low temperatures a liquid A-liquid B interface is formed.}
\label{den-profiles-a}
\end{center}
\end{figure}

\begin{figure}
\begin{center}
\includegraphics[height = 5.0in, width=5.0in, keepaspectratio=true]
{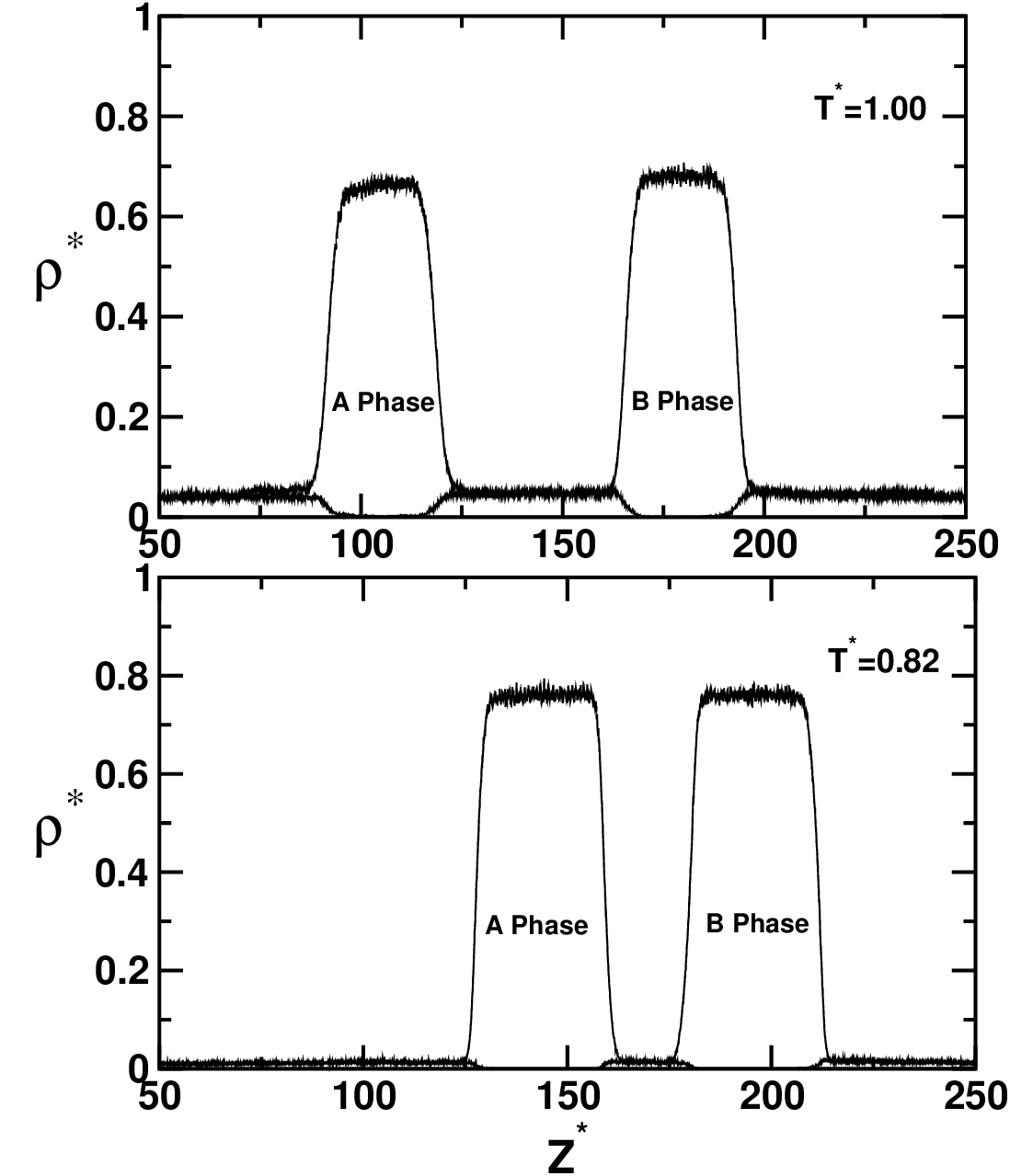}
\caption{Liquid-vapor-liquid reduced density profiles of the mixture
for $ \alpha=0.25 $  and $N=4096$. At these higher temperatures 
the vapor phase spontaneously wets the liquid A-liquid-B  phases.}
\label{den-profiles-b}
\end{center}
\end{figure}

\begin{figure}
\begin{center}
\includegraphics[height = 5.0in, width=5.0in, keepaspectratio=true]
{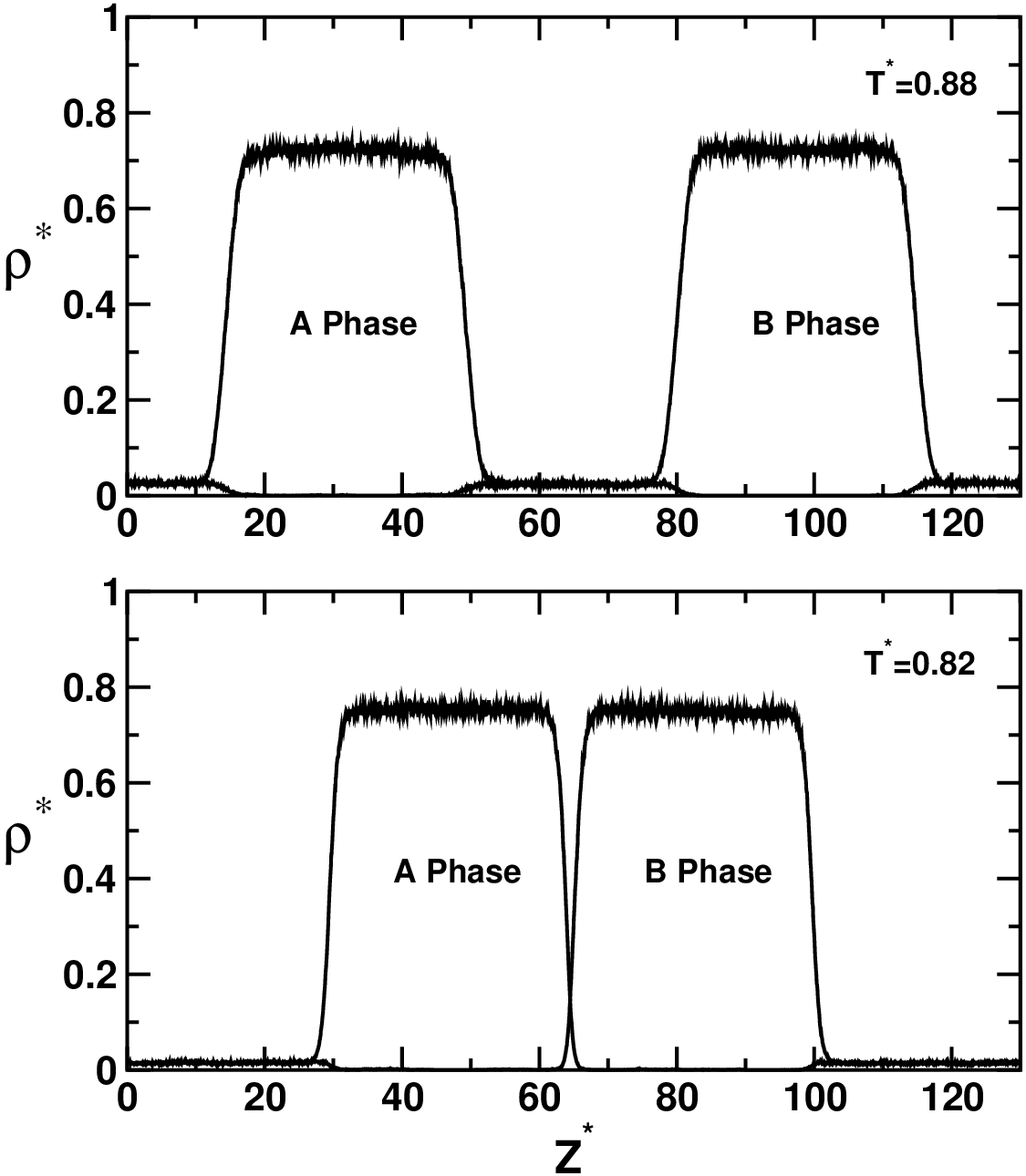}
\caption{Liquid-vapor-liquid reduced density profiles of the mixture
for $ \alpha=0.30 $  and $N=4096$. At these higher temperatures 
the vapor spontaneously wets the liquid A-liquid-B  phases.}
\label{den-profiles-c}
\end{center}
\end{figure}

\begin{figure}
\begin{center}
\includegraphics[height = 5.0in, width=5.0in, keepaspectratio=true]
{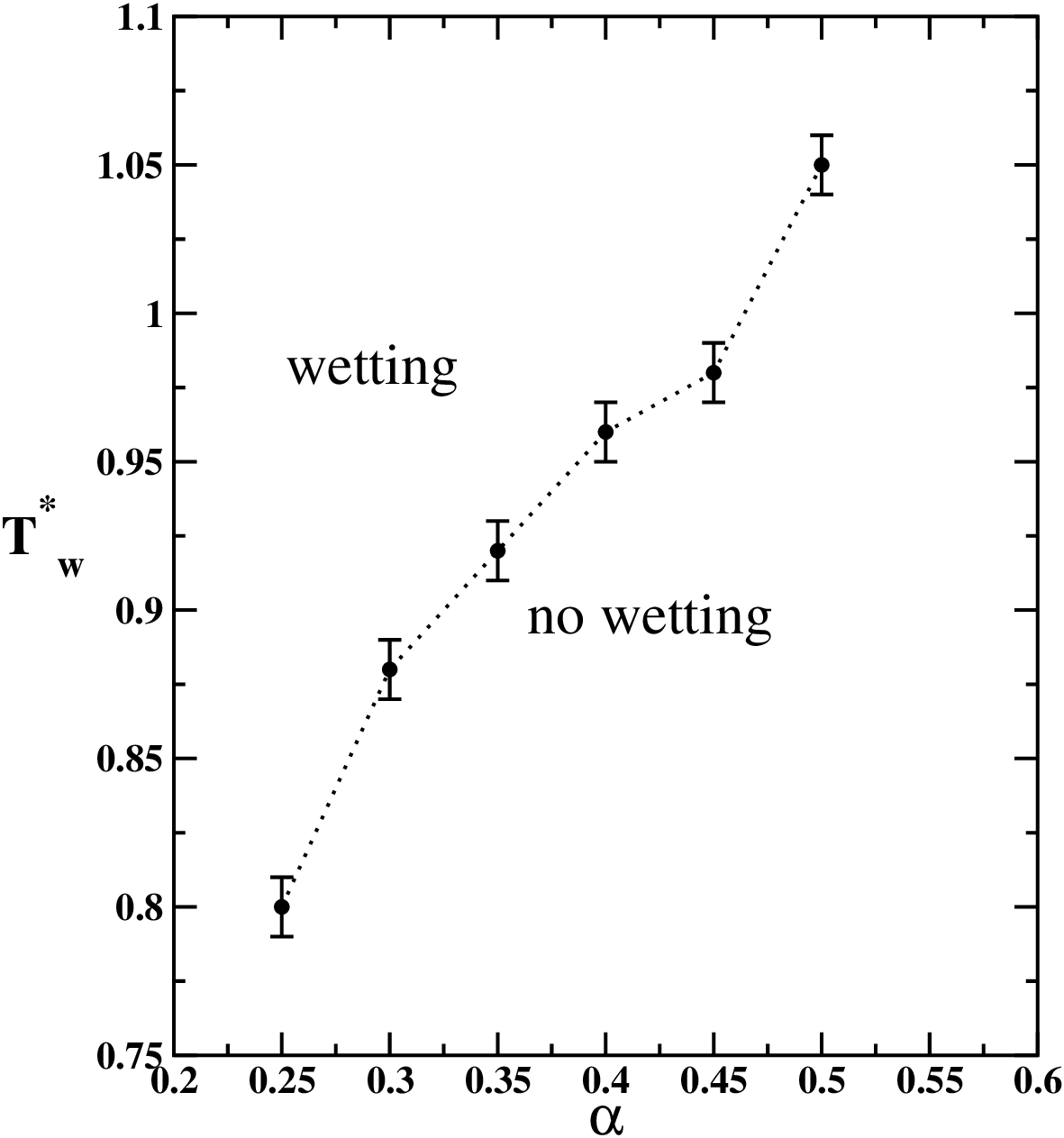}
\caption{Reduced wetting transition temperature as function
of $ \alpha $ for an equimolar binary mixture. These results were obtained 
from MD simulations with a $N=4096$ particles. The line is a guide 
to the eye.}
\label{Tw-alpha}
\end{center}
\end{figure}

Once the system reached equilibrium we calculated the structural
properties of the system from simulations with $N=4096$ molecules. 
In figure~\ref{den-profiles-a} we show the density profiles at the 
relatively low temperatures, $ T^{*}=0.65 $ and $0.75$, when $ \alpha=0.25 $.
In this temperature region the liquid-vapor equilibrium structure of the 
mixture consists of a liquid A-liquid B interface in coexistence with the 
vapor phase.  As temperature increases and reaches the region 
$ 0.80 \leq T^{*} \leq 1.25$, however, this fluid phase structure 
rearranges  in such a way that, the vapor phase spontaneously sets 
in between the Liquid A-Liquid B phases, as shown in 
figure~\ref{den-profiles-b}. This structure remains stable during all 
the time of the simulation, about 55 ns in the scale
of argon. This is a clear evidence that {\it the vapor phase wets 
the liquid phases}. In figure~\ref{den-profiles-c} we show similar
results for $ \alpha=0.3 $. There one sees that at, $ T^{*} =0.82$, the 
structure of the system is such that there is a liquid A-liquid B interface.
Nonetheless, at higher temperatures, for instance, $ T^{*}=0.9 $, the vapor 
phase wets the liquid phases. This behavior of $ \rho^{*}(z) $ for 
$ \alpha=0.3 $, suggest that 
$ T_{\rm W}^{*}(\alpha=0.25) < T_{\rm W}^{*}(\alpha=0.3) $. 
In fact, comparing the structures plotted in figures~\ref{den-profiles-b}
and \ref{den-profiles-c}, one  should note that the mixture with 
$ \alpha=0.25 $ already wets at, $ T^{*} =0.83$. 
Following a systematic analysis of the structure of density profiles 
as function of temperatures for all the values of $ \alpha $ we estimated 
the wetting transition temperatures $ T_{\rm W}^{*}( \alpha) $. We found 
that $ T_{\rm W}^{*} $ increases monotonically as a function of $\alpha$, 
whenever $ 0< \alpha \leq 0.5 $. For higher values,  $0.5 <  \alpha < 1.0$,
this wetting phenomenon does not occur. The results are summarized in the 
wetting phase diagram in figure~\ref{Tw-alpha}. 
We believe that the reason for which the system 
wets below $ \alpha=0.5 $, and no longer does above  this value is  
due to the equal size of  the molecules of type A and B and because
$ \varepsilon_{AA}=\varepsilon_{BB} $.

\begin{figure}
\begin{center}
\includegraphics[height = 5.0in, width=5.0in, keepaspectratio=true]
{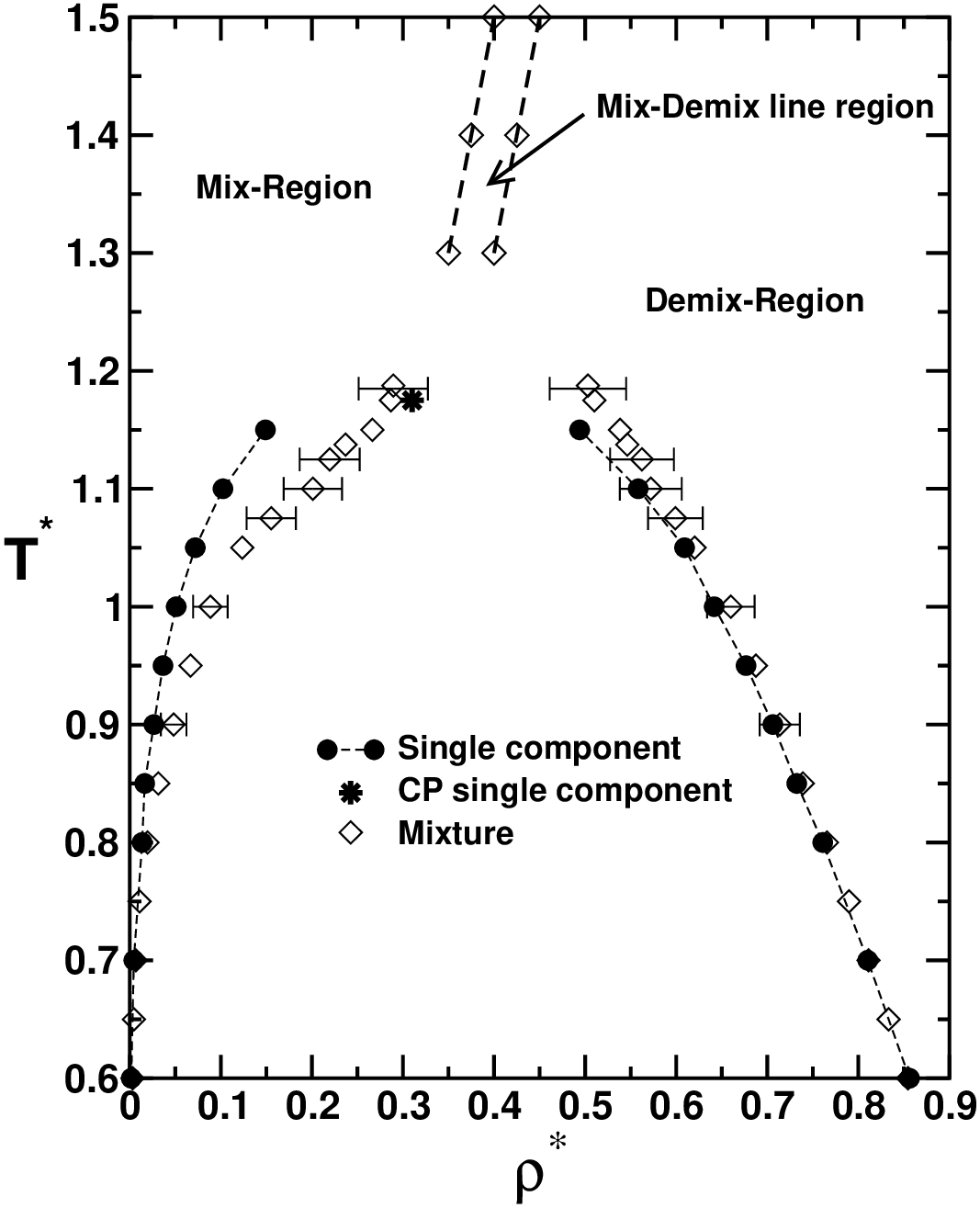}
\caption{$ T^{*} $ versus $ \rho^{*} $ phase diagram for a partially 
miscible mixture with $ N=4096 $ and $ \alpha=0.25 $. For comparison
we have included the corresponding phase diagram of a single LJ
fluid.}
\label{Phase-diag}
\end{center}
\end{figure}

Now we will try to relate these density profiles 
structure and wetting phenomenon with the properties of the phase diagram of 
the mixture.  To this end we will discuss in detail the properties of the corresponding 
phase diagram for $ \alpha=0.25 $. To begin it is important to remind 
that the present model binary mixture corresponds to the type III in the 
classification of Scott and Konynenburg~\cite{scott70}. The phase diagram 
properties of this kind of mixtures has been quantitatively studied by Wilding 
{\it et. al.}~\cite{wilding98}, using a square well  potential for the 
intermolecular interactions. They showed that for a strong immiscible binary 
mixture a tricritical point exists. This means that the $\lambda$ line meets 
the LV coexistence curve at the critical point. In figure~\ref{Phase-diag} we 
show the $T^{*}$ versus $\rho^{*}$ phase diagram of the mixture obtained from 
extensive MD simulations with $ \alpha=0.25 $. Since for $ \alpha=0 $ and 1 our 
model reduces to single LJ fluids, we have also included in the same figure,
for comparison, the phase diagram of a single LJ fluid. 
Both phase diagrams were calculated simulating a system with $ N=4096 $ molecules 
and using a shifted intermolecular potential with a cutoff of $3\sigma$. 
Therefore, the LV critical temperature 
of the single fluid became $ T_{C}^{LJ} \simeq 1.2$.  
One should note that, the mixture critical density is higher than that 
of the single LJ fluid, and its critical temperature shifts upwards.
This shift in $ \rho^{*} _{c}$ and $ T^{*}_{c} $ occurs due to the fact that  
the less miscible the mixture is, --smaller $ \alpha $--, the larger the 
temperature range of immiscibility. In addition, the  $\lambda$ line appears to
touch the LV critical point, and  the critical point becomes 
tricritical. These results agree well with those obtained recently 
for square well strong immiscible binary mixtures~\cite{wilding98}.  

\begin{figure}
\begin{center}
\includegraphics[height = 5.0in, width=5.0in, keepaspectratio=true]
{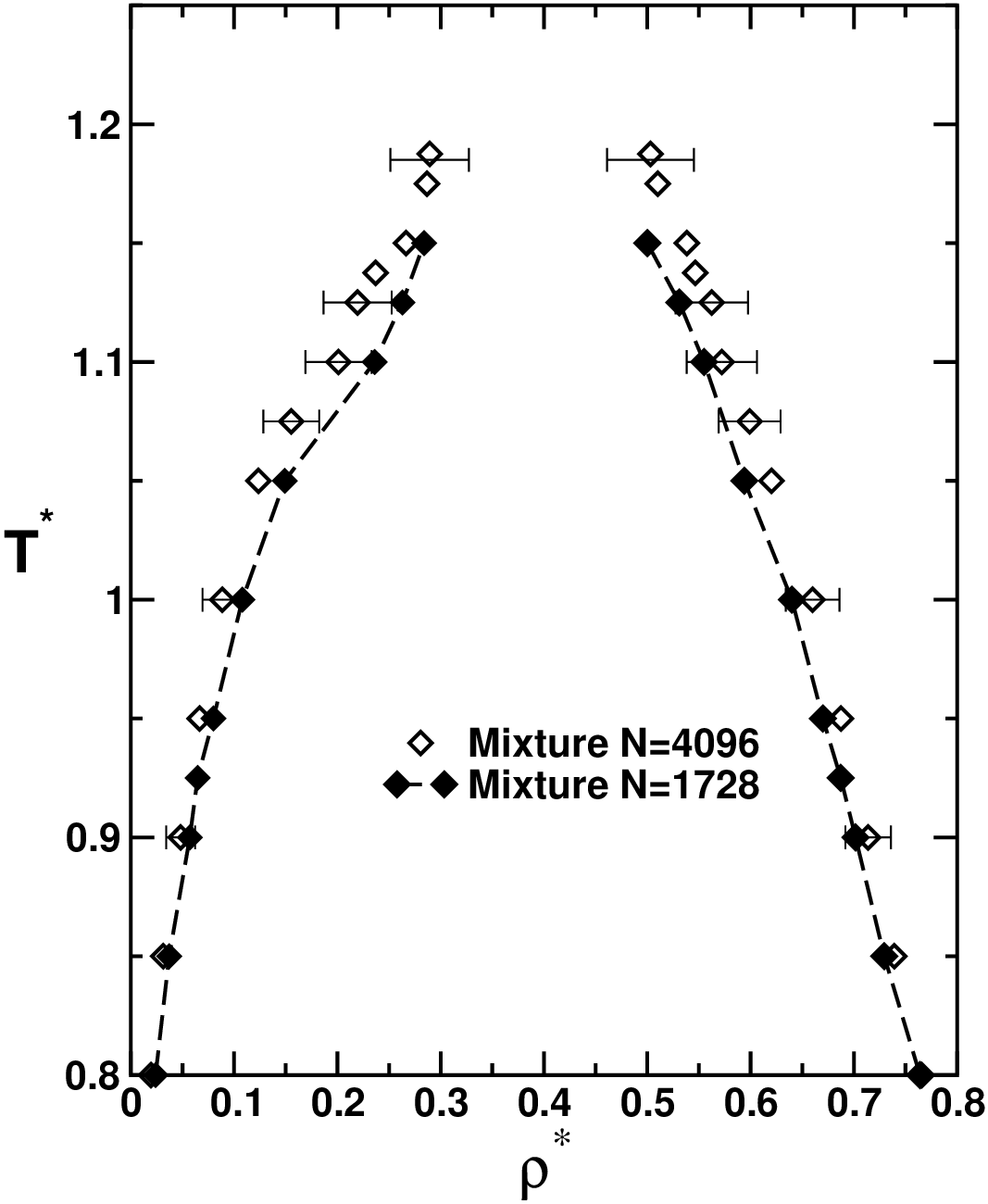}
\caption{$ T^{*} $ versus $ \rho^{*} $ phase diagram for a binary 
mixture with $ N=1728 $ ( $\blacklozenge$ ) and 
$ N=4096 $  ($\diamondsuit$)  particles and $ \alpha=0.25 $.}
\label{tamanio}
\end{center}
\end{figure}

\begin{figure}
\begin{center}
\includegraphics[height = 5.0in, width=5.0in, keepaspectratio=true]
{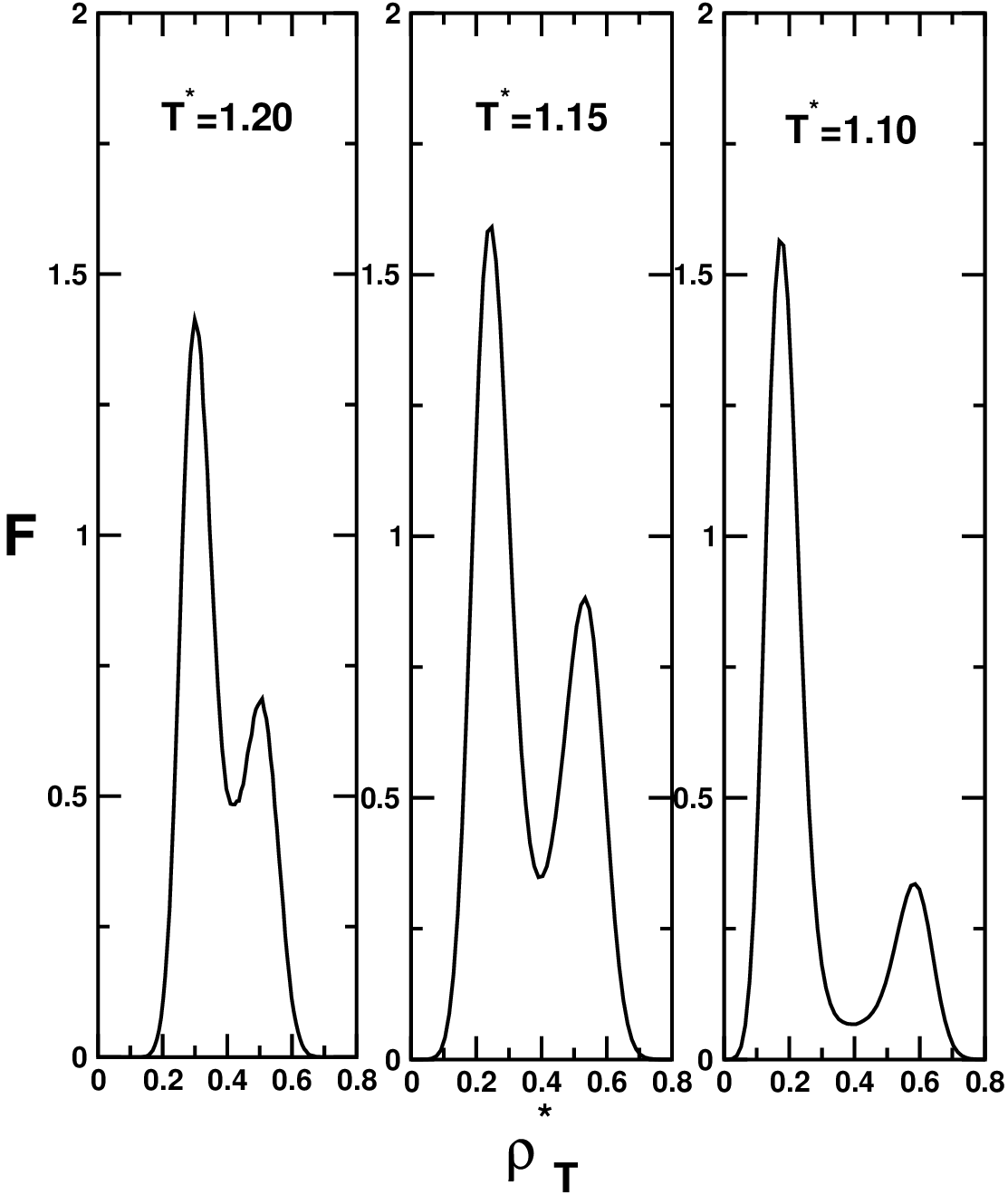}
\caption{Frequency versus total reduced density (total density distribution)
of the mixture with $ \alpha=0.25 $ and $N=4096$ at $ T^{*}=1.1, 1.15 $,
and 1.2.  The vertical axis should be multiplied by a factor of $ 10^{3} $.
These temperatures are slightly higher than  the tricritical point temperature.}
\label{histo-density}
\end{center}
\end{figure}

To check how sensitive are the phase diagram properties of the mixture,
for $ \alpha =0.25 $, to the number of molecules in the simulations, 
we also calculated the phase boundaries simulating a system with $N=1728$ 
molecules. In fig.~\ref{tamanio} we show the results  of this finite size 
analysis. For a mixture with $N=1728$ the density of the liquid phase 
decreases while the density of the vapor phase increases. We also observe 
that at low temperatures the results are  system size independent.
However, as temperature increases, in particular, close to 
the critical point, there are differences in the phase boundaries obtained
using $ N=1728 $ and 4096 molecules.  As one approaches the critical point  
it turns out  more difficult to determine the coexistence densities with
a system with $N=1728$ molecules. This is so because the difference in the 
coexistence densities  becomes smaller, and the number of particles in the 
system is not sufficient to give rise to bulk fluid phases. Nevertheless, 
for a system with $N=4096$ this is not the case and we indeed obtained  
the liquid and vapor fluid phases. On the other hand, it is known that near 
to the critical point the fluctuations of the density are strong and the 
vapor and liquid densities are not well defined. 
This fact complicates the location of the critical point. To circumvent
this difficulty we proceed as in reference~\cite{binder} and 
calculated the total density distributions in a system
with $N=4096$, at several temperatures around $ T_{c} $. 
The simulational box was divided in several slabs, parallel to the interface,
 of width between $ \sigma$  and $ 3\sigma $ . 
The  density of particles was calculated in each slab every 50 
time steps of the MD simulations.
A block average histogram of the densities is obtained every 
$ 5\times 10^{4} $  time steps. The resulting total density distribution,
$\rho_{A} + \rho_{B}  $, was 
calculated after averaging over 20 blocks. The  result  is presented in 
figure~\ref{histo-density} .  At temperatures 
slightly below $ T_{c} $, we obtained density distributions that show two 
maxima. The low density maximum corresponds to the vapor phase and the 
higher maximum corresponds to the liquid density. Nonetheless, at temperatures 
above $ T_{c} $, the density distributions show only one maximum. 
In Fig~\ref{histo-density} we show the total density distributions at 
$T^{*}=1.1,\>1.15$ and $1.2$. The low density maximum is higher because 
the volume of the vapor phase is larger than the volume of the liquid phases. 
Thus, the vapor density appears with a higher frequency in the histograms.  

\begin{figure}
\begin{center}
\includegraphics[height = 6.0in, width=6.0in, keepaspectratio=true]
{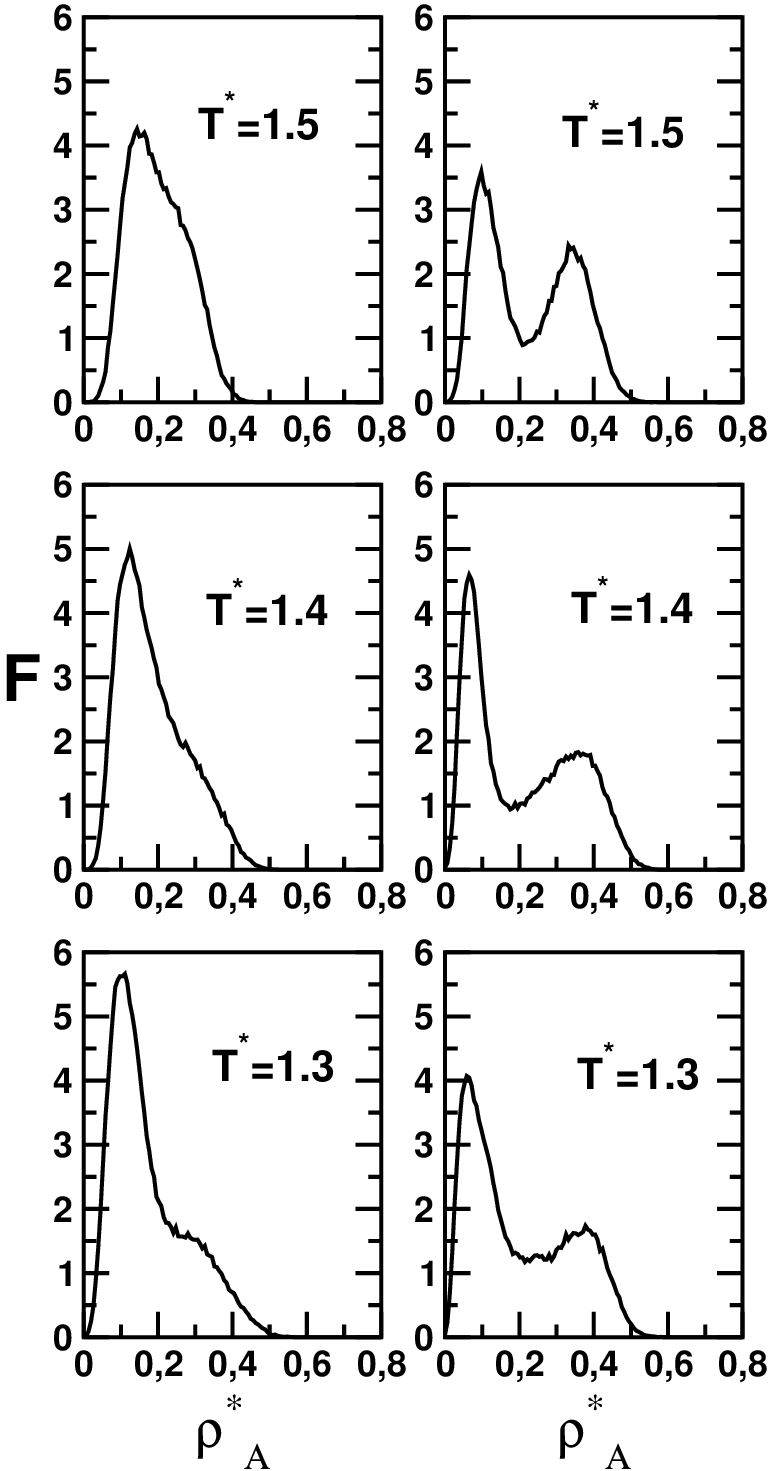}
\caption{Frequency versus reduced density (density distribution)
for one of the fluid phases (fluid A) with $ \alpha=0.25 $ and $N=4096$. 
The vertical axis should be multiplied by a factor of $ 10^{3} $.
The left columns, show the density distributions when the $\lambda$-line 
is approached form the mixing region at $T^{*}=1.3, 1.4$, and 1.5. 
Note that they show a one peak structure. However, 
if the $\lambda$-line is approached from the demixing region at $T^{*}=1.3, 1.4$, 
and 1.5 the density distributions, right columns, show a two peak structure. }
\label{histo-demix}
\end{center}
\end{figure}

To locate of mixing-demixing transition temperatures, --$ \lambda $ line-- 
for $ \alpha=0.25 $ and $N=4096$, we followed a similar procedure as that 
described for  the location of the LV critical point. However, in the 
determination of the $\lambda$ line, we only consider the density 
distribution of one of the species, $\rho_A$ or $\rho_B$. 
The reason is that when the system 
is in the demixing region, the density distribution of one of the species 
shows two  maxima. The low density peak corresponds to the poor fluid phase 
and the high density peak correspond to the rich fluid phase. Nonetheless, 
when the system is in the mixing region the fluid phases become homogeneous 
and therefore one should observe only one peak in the density distribution. 
Therefore, to locate the mixing-demixing points we looked at the transition 
from the two peak structure density distribution to one peak density distribution. 
This analysis was done as a function of the total density of the system and at 
three different temperatures. The results are shown in figure~\ref{histo-demix}.

\begin{figure}
\begin{center}
\includegraphics[height = 5.0in, width=5.0in, keepaspectratio=true]
{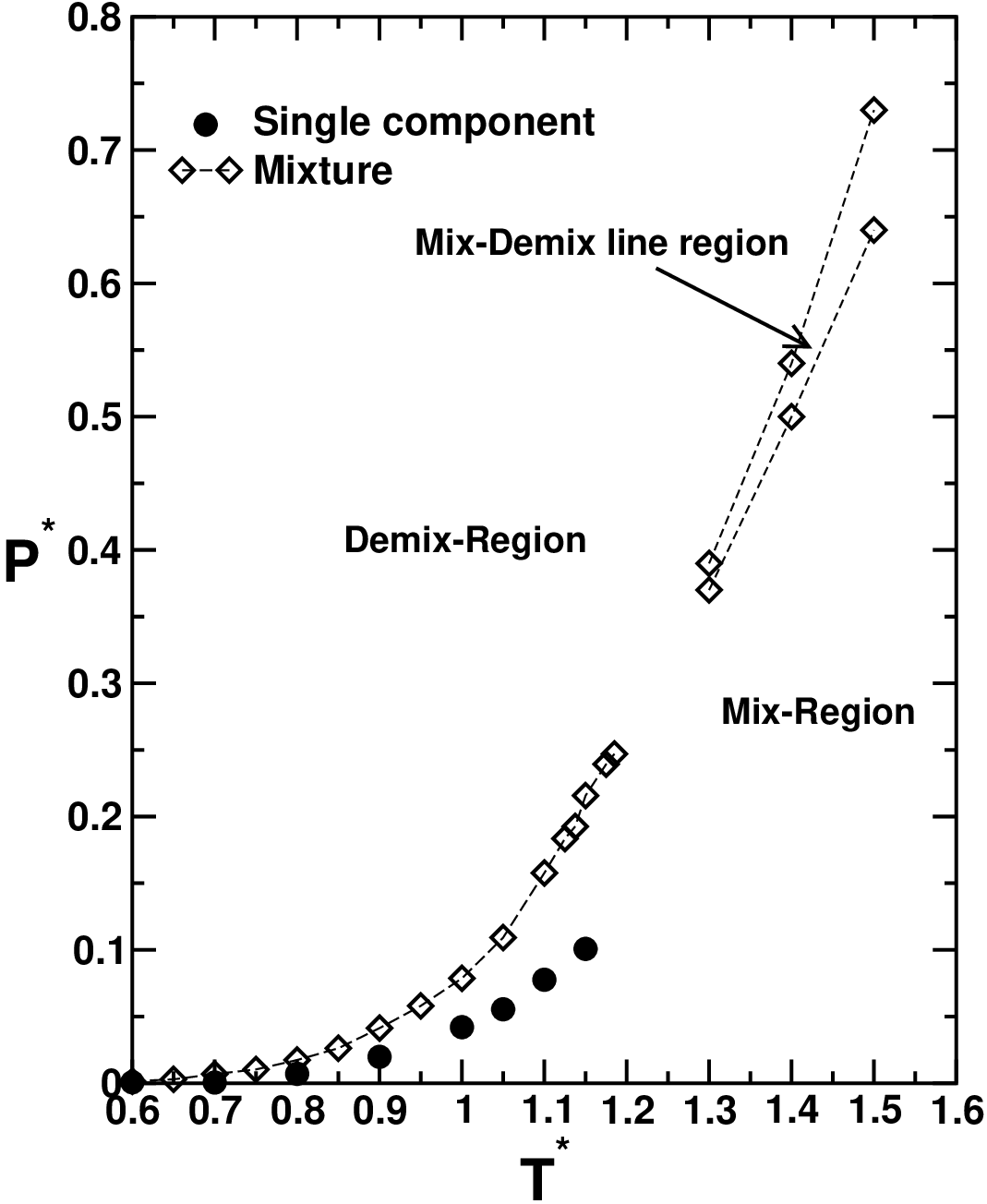}
\caption{Reduced pressure versus reduced temperature 
phase diagram for  $ \alpha=0.25 $.}
\label{press-temp}
\end{center}
\end{figure}

Another way of locating   the $\lambda$ line  
is calculating the pressure versus temperature phase diagram, shown in 
figure~\ref{press-temp}, for a system with $N=4096$ molecules.  The 
pressure was calculated  as the average of the pressure tensor component 
perpendicular to the interface via the virial formula~\cite{rowlin-capillary}.
There we also included for comparison, the results for a single LJ fluid with the 
same number of molecules. We found that the LV phase boundary is located 
right at $ T^{*} \simeq 1.25$.  At higher temperatures there are 
two branches that were obtained by approaching the mixing-demixing boundary 
from both sides of the  $ \lambda $ line. These branches enclose a region 
that is narrower in size, as compared to the region obtained in the 
$\rho^{*}$ versus $T^{*}$ phase diagram. These results suggest that 
the calculation of the $ P^{*} $ versus $ T^{*} $ phase diagram gives
more precise way to locate the mixing-demixing line.
Again, we did find  evidence that suggest that the $\lambda$ line meets the LV 
coexistence line at the LV critical point. These results give a strong 
support to the existence of  a tricritical point~\cite{wilding98}. 
\subsection{Interfacial properties}
\label{interfacial}
\begin{figure}
\begin{center}
\includegraphics[height = 5.0in, width=5.0in, keepaspectratio=true]
{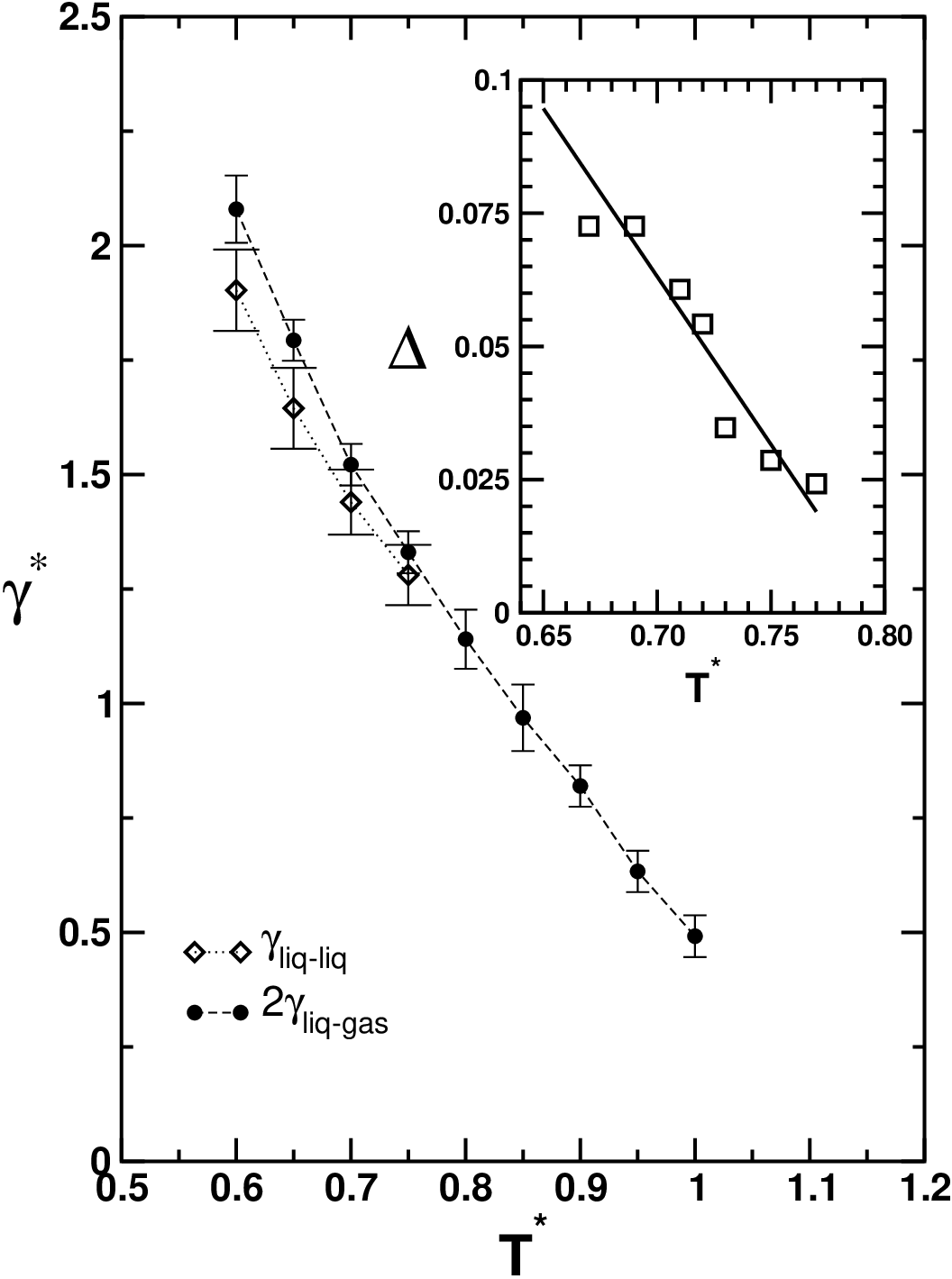}
\caption{Reduced interfacial tension as function of reduced temperature 
for both, LL  and LV interfaces.  In the inset we plot the difference 
$ \Delta =2\gamma_{\rm LV} - \gamma_{\rm LL}$ as a function 
of temperature. These results are representative of a mixture with 
$ \alpha=0.25 $.}
\label{surf-tension}
\end{center}
\end{figure}

\begin{figure}
\begin{center}
\includegraphics[height = 5.0in, width=5.0in, keepaspectratio=true]
{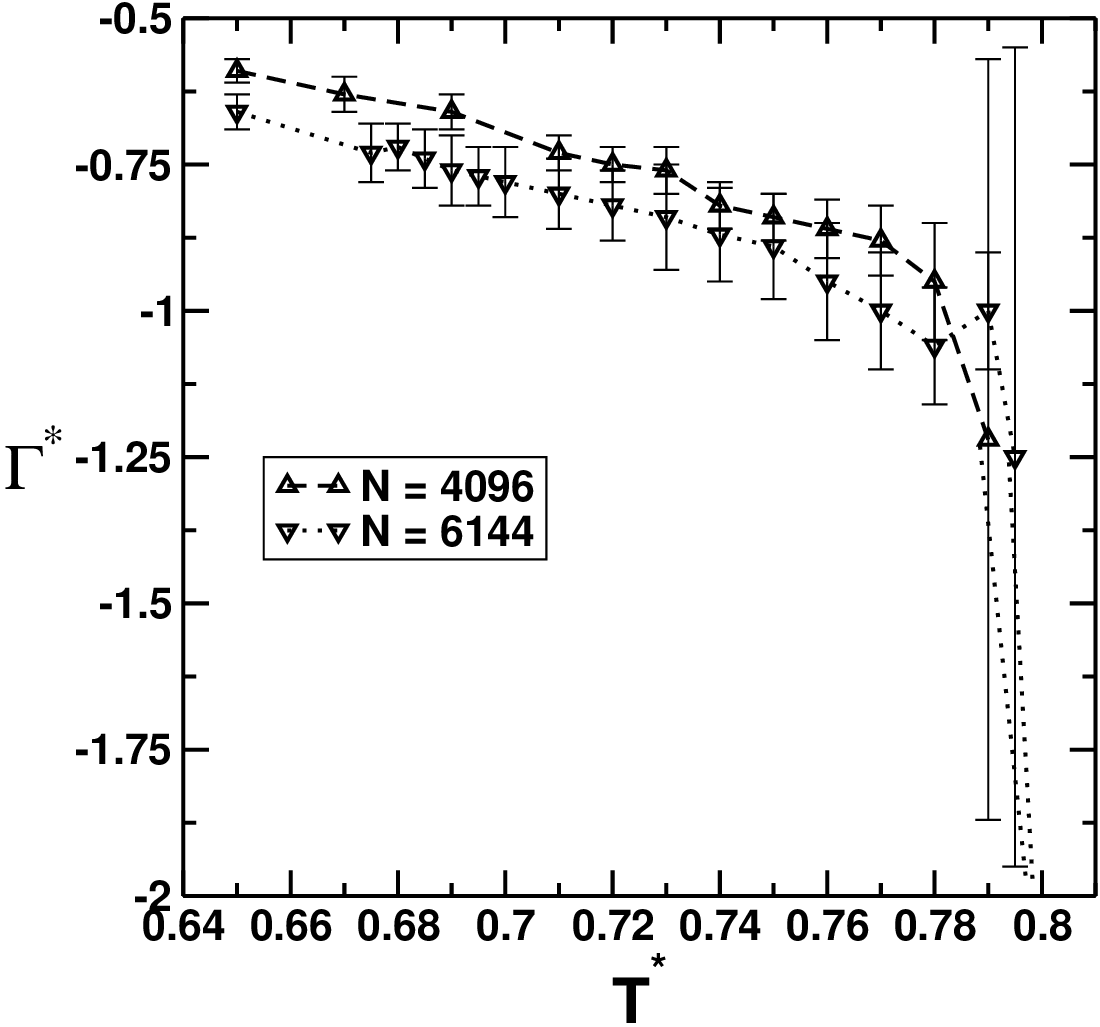}
\caption{Reduced adsorption as function of reduced temperature
at the LL interface for two systems, N=4096 ($\bigtriangleup$), 
N=6144 ($\bigtriangledown$) and $ \alpha=0.25 $.
The two points that are closer  to the wetting transition 
temperature show relatively large error bars. This is due to 
the increase of fluctuations of the interface width close to 
$ T^{*}_{\rm W} (\alpha) $. In the present case,
the vapor wets the LL interface at, $T^{*}_{\rm W}(\alpha)=0.80$,
and the adsorption jumps from a finite negative value up to minus 
infinite. Dashed lines are a guide to the eye.}
\label{adsorption}
\end{center}
\end{figure}

Now we turn to the discussion of the interfacial properties and
surface phenomenon of the mixture at the liquid-vapor phase 
coexistence.  We carried out a quantitative analysis of 
these properties for $ \alpha=0.25 $. To evaluate the wetting transition 
temperature with precision we calculated the interfacial free energies 
as a function of temperature. To this purpose we use the well known formula,
\begin{equation}
\gamma = \int^{\rm bulk_2}_{\rm bulk_1} 
\Big(P_n(z)-P_t(z)\Big)dz,
\label{gamma}
\end{equation}
where the integrations were carried out up to the middle of the 
corresponding bulk phases. The tangential and normal pressure 
profiles were calculated using the definition of the 
Irving-Kirkwood pressure tensor~\cite{rowlin-capillary}. 
For a planar interface these pressure
profiles are given by the formula~\cite{rowlin-capillary,rao}. 
\begin{eqnarray}
P_{n}(z)&=&\rho(z)k_{B}T  \\
&-& \frac{1}{2A} \Big\langle \sum_{i\neq j}
\frac{z_{ij}^{2}u'_{ij}(r_{ij})}{r_{ij}|z_{ij}|} 
\theta\Big(\frac{z-z_{i}}{z_{ij}}\Big) \theta\Big(\frac{z_{j}-z}{z_{ij}}\Big)
\Big\rangle,\nonumber
\label{pressure}
\end{eqnarray} 
\begin{eqnarray}
P_{t}(z)&=&\rho(z)k_{B}T \\ \nonumber
&-& \frac{1}{4A} \Big\langle \sum_{i\neq j}
\frac{[x_{ij}^{2}+ y_{ij}^{2}] u'_{ij}(r_{ij})}{r_{ij}|z_{ij}|} 
\theta\Big(\frac{z-z_{i}}{z_{ij}}\Big) \theta\Big(\frac{z_{j}-z}{z_{ij}}\Big)
\Big\rangle.
\end{eqnarray} 
According to  Young's rule the difference,
\begin{equation}
\Delta (T)=2\gamma_{\rm LV} - \gamma_{\rm LL},
\end{equation}  
must be zero at the WTT. So, the wetting by the 
vapor phase occurs when the free energy difference, $\Delta$, 
becomes negative. This quantity is plotted as 
a function of reduced temperature in the inset of 
figure~\ref{surf-tension} for $ \alpha=0.25 $ and $ N=4096 $. 
A linear extrapolation of the data indicates that the 
wetting occurs at about $T_{\rm W}^{*}(\alpha=0.25 )=0.80$.  Notice that
due to the planar geometry of the interfaces there is no contact angle 
and the three surface tensions are independent and considered separately.
To improve the accuracy of $ T_{\rm W}^{*} (\alpha=0.25 )$, one needs to 
carry out even more demanding simulations. This is due to the fact that 
the interfacial tension always shows relatively large fluctuations. The
situation complicates even more when the simulations are performed 
at temperatures very close to the WTT. 
A second independent way to estimate the WTT and to figure out the nature 
of the WT, we calculated the adsorption of molecules at the LL interfaces
shown in figures~\ref{den-profiles-a} . 
This is done using the formula,
\begin{equation}   
 \Gamma =\int^{\rm bulk B}_{\rm bulk A} 
\Big( \rho(z) - \rho_{\rm bulk}\Big) dz,
\end{equation} 
The results of the calculations for $ \alpha=0.25 $ and $N=4096$ and $6148$
are plotted in figure~\ref{adsorption}. One sees that 
$ \Gamma(T^{*},\alpha=0.25)$ 
decreases monotonically in the temperature range, 
$ 0.75 \leq T^{*} \leq 0.78$, 
and it consistently shows  negative values. This is due to the 
inhomogeneity at the LL interface, since the density there, is much 
smaller than the density of the liquid bulk phases. Nonetheless, as  
$T^{*}\rightarrow 0.80 $ from below the adsorption jumps from a finite 
negative value up to minus infinite since the vapor wets the LL interface at 
$ T^{*}(\alpha=0.25)\simeq 0.80 $. Note that the WTT shifts slightly towards 
higher temperatures as the number of molecules in the system increases from 
$N=4096$ up to $N=6148$. This {\it jump or discontinuity } is  a strong  
indication that the WT is of first order. As expected the closer the 
temperature approach to $T^{*}_{W}(\alpha)$  the stronger the fluctuations 
in $ \Gamma(T^{*},\alpha)$. 
In figure~\ref{adsorption} the closest approach to the WTT, in reduced 
temperature, was $ \frac{\delta T}{T_{W}} =6.25 \times 10^{-3}$.
One would expect that  the wetting transition is of first order even
for other values of the miscibility parameter, $ 0<\alpha\leq 0.5 $. This is
so,  since the behavior of $ \rho^{*} (z, \alpha)$ is similar to that of $ \alpha=0.25 $,
whose surface properties  were studied in detail.
\section{Conclusions}
\label{conclusions}
We carried out extensive MD simulations to study 
the LV phase coexistence, the structural properties 
and interfacial phenomena of a partially 
miscible symmetrical LJ binary mixture. By analyzing the density
profiles as a function of temperature and miscibility parameter
we estimated the wetting phase diagram, $ T_{\rm W}^{*} $ versus
$ \alpha $.  The wetting of the vapor phase happens whenever
$ 0 <  \alpha  \leq 0.5$. We also found that $ T^{*}_{\rm W} (\alpha)$,
monotonically increases  up to $ \alpha=0.5 $. For other values of 
$ \alpha $, this wetting phenomenon does not occur. In addition, 
we also studied quantitatively the $ T^{*} $ versus $ \rho^{*} $  
and $ P^{*} $ versus $ T^{*} $ phase diagrams for $ \alpha=0.25 $.  
The results indicate that the former phase diagram shows a similar 
topology as that obtained for a square-well potential mixture estimated 
by means of mean-field theory and Monte Carlo simulations~\cite{wilding98}.  
An analysis of the behavior of the adsorption of
particles, at the LL interface, as a function of $ T^{*} $,  
for $ \alpha=0.25 $,
led to the conclusion that the WT is of first order. 
These results should be valid for a family of mixtures of the type III,
in the classification of Scott and Konynemburg. 
The phase diagrams discussed here together with the wetting phenomenon are an 
explicit quantitative demonstration  of the  scenario suggested some time ago 
based on a microscopic expression for the Hamaker constant~\cite{getta93}.
To the best of our knowledge {\it this is the first time that this wetting 
phenomenon  is  quantitatively  studied by means of MD simulations
in binary LJ fluid mixtures in absence of a wall.}. We would like to point 
out that this wetting phenomena agrees well  with some experimental studies 
carried out with fluid binary mixtures. Finally, the results reported in this 
paper provide with a more complete understanding of the surface phenomena in 
partially miscible fluid binary mixtures.
\section{Acknowledgments}  
EDH would like to thank the DAAD-Germany for 
financial support during the summer of 1999. Partial support 
from CONACYT contracts L0080-E9607 (EDH) and 25298-E (GRS) 
as well as from DGAPA-UNAM contracts IN110103, and IX107704, 
is also acknowledge.

\newpage
\newpage
\newpage
\newpage
\newpage
\newpage
\vspace{4in}
\newpage
\vspace{4in}
\newpage
\vspace{4in}
\newpage 
\newpage
\newpage
\newpage

\end{document}